\documentclass{llncs}
\usepackage{graphicx}
\usepackage{amsmath}
\usepackage{amssymb}
\usepackage{enumerate}
\usepackage{varioref}

\def\boxit#1{\vbox{\hrule\hbox{\vrule\kern4pt
  \vbox{\kern1pt#1\kern1pt}
\kern2pt\vrule}\hrule}}

\newcommand{\es}{\varnothing}

\title{New Parameterized Algorithms for Edge Dominating Set}

\author{Mingyu Xiao\inst{1} \and
Ton Kloks\inst{2} \and
Sheung-Hung Poon\inst{2}
}

\institute{
 School of Computer Science and Engineering\\
 University of Electronic Science and Technology of China\\
China\\
{\tt myxiao@gmail.com}
\and
Department of Computer Science\\
National TsingHua University\\
Taiwan\\
{\tt \{kloks,spoon\}@cs.nthu.edu.tw}}

\begin{document}

\maketitle

\begin{abstract}
An edge dominating set of a graph $G=(V,E)$
is a subset $M \subseteq E$ of edges in the graph such
that each edge in $E-M$ is incident with at least one edge in $M$.
In an instance of the parameterized edge dominating set problem
we are given a graph $G=(V,E)$ and an integer $k$
and we are asked to decide whether $G$ has an edge dominating
set of size at most $k$.
In this paper we show that the parameterized edge dominating
set problem can be solved in $O^*(2.3147^k)$ time and polynomial
space.
We show that this problem can be reduced to
a quadratic kernel with $O(k^3)$ edges.
\end{abstract}

\section{Introduction}

The {\em edge dominating set problem} (EDS),
to find an edge dominating set of minimum size in a graph,
is one of the basic problems highlighted by Garey and Johnson
in their work on NP-completeness~\cite{kn:garey}.
It is known that the problem is NP-hard even when
the graph is restricted to planar or bipartite graphs
of maximal degree three~\cite{kn:yannakakis}.
The problem in general graphs and in sparse graphs has been extensively
studied in approximation
algorithms~\cite{kn:yannakakis,kn:fujito,kn:cardinal}.
Note that a maximum matching is a 2-approximation for EDS.
The 2-approximation algorithm for the weighted version of EDS
is considerably more complicated~\cite{kn:fujito}.

\smallskip
Recently, EDS also draws much attention from the exact - and
parameterized algorithms community.
Randerath and Schiermeyer~\cite{kn:randerath}
designed an $O^*(1.4423^m)$ algorithm for EDS
which was improved to  $O^*(1.4423^n)$ by
Raman {\em et al.}~\cite{kn:raman}.\footnote{The $O^*$-notation
suppresses polynomial factors.} Here
$n$ and $m$ are the number of vertices and edges in the graph.
Fomin {\em et al.},~\cite{kn:fomin}
further improved this result to $O^*(1.4082^n)$ by
considering the treewidth
of the graph.
Rooij and Bodlaender~\cite{kn:rooij} designed an $O^*(1.3226^n)$
algorithm by using the `measure and conquer method.'

\smallskip
For \emph{parameterized edge dominating set} (PEDS) with the parameter $k$
being the size of the edge dominating set, Fernau~\cite{kn:fernau}
gives an $O^*(2.6181^k)$ algorithm.
Fomin {\em et al.}~\cite{kn:fomin} obtain
an $O^*(2.4181^k)$-time and exponential-space
algorithm based on dynamic programming on bounded treewidth graphs.
Unfortunately, their paper only briefly sketches the
description and analysis of this algorithm.

\smallskip
Faster algorithms are known for graphs that have maximal degree three.
The EDS and PEDS problems in degree-3 graphs can be solved
in $O^*(1.2721^n)$~\cite{kn:xiao2}
and  $O^*(2.1479^k)$~\cite{kn:xiao3}.

\smallskip
In this paper, we present two new algorithms for PEDS.
The first one is a simple and elegant algorithm
that runs in $O^*(2.3715^k)$ time and
polynomial space.
We improve the running-time bound
to $O^*(2.3147^k)$ by using a technique that deals with remaining graphs
of maximal degree three.
We also design a linear-time algorithm that
obtains a quadratic kernel which is smaller than previously-known kernels.

Our algorithms for PEDS are based on the technique of enumerating
minimal vertex covers. We introduce the idea of the
algorithms in Section~\ref{sec_enumeration} and
introduce some basic techniques in Section~\ref{sec_rules}.
We present a simple algorithm for PEDS in Section~\ref{sec_firsalg}
and an improved algorithm in Section~\ref{sec_improved}.
We moved the proof of a
technical lemma to Appendix~\ref{analy3}.
In Section~\ref{sec_kernel} we discuss the problem kernel.

\section{Enumeration-based algorithms}
\label{sec_enumeration}

As in many previous algorithms for the edge
dominating set
problem~\cite{kn:fernau,kn:fomin,kn:rooij,kn:xiao2}
our algorithms are based on the enumeration of minimal vertex covers.
Note that the vertex set of an edge dominating set is a vertex cover.
Conversely, let $C$ be a minimal vertex cover and $M$ be a minimum edge dominating set containing $C$ in the set of its endpoints. Given $C$, $M$
can be computed in polynomial time by computing a maximum matching in induced graph $G[C]$ and adding an edge for each unmatched vertex in $C$.
This observation reduces the problem to that of finding the right
minimal vertex cover $C$.
Now, the idea is to enumerate all minimal vertex covers.
Moon and Moser showed that the number of minimal vertex covers is
bounded by $3^{n/3}$ and this shows that
one can solve EDS in $O(1.4423^n)$
time~\cite{kn:johnson,kn:moon}.

For PEDS, we want to find an edge dominating set of size bounded by $k$.
It follows that we need to enumerate minimal vertex covers of size
only up to $2k$. We use a branch-and-reduce method to find vertex covers.
We fix some part of a minimal
vertex cover and then we try to extend it with
at most $p$ vertices. Initially $p=2k$. In fact, in our algorithms, we may not really enumerate all minimal vertex covers of size up to $p$. But we will guarantee that at least one of the right vertex covers will be considered if a solution exists.

\smallskip
For a subset $C\subseteq V$ and an independent set
$I\subseteq V-C$ in $G$,
an edge dominating set $M$ is called a {\em $(C,I)$-eds}
if
\[C\subseteq V(M) \quad\text{and}\quad I\cap V(M)=\es.\]
In the search for the vertex cover $V(M)$ of a minimum $(C,I)$-eds $M$,
we keep track of a partition of the vertices of $G$ in four sets:
$C$, $I$, $U_1$ and $U_2$.
Initially $C=I=U_1=\es$ and $U_2=V$.
The following conditions are kept invariant.
\begin{enumerate}
\item
$I$ is an independent set in $G$, and
\item
each component of $G[U_1]$
 is a clique component of $G[V \setminus (C\cup I)]$.
\end{enumerate}
The vertices in $U_1\cup U_2$ are called {\em undecided\/} vertices.
We use a five-tuple
\[(G,C,I,U_1,U_2)\]
to denote the state described above.
We let $q_i=|Q_i|$ denote the number of vertices of a clique component
$Q_i$ of $G[U_1]$.
Rooij and Bodlaender proved the following
lemma in~\cite{kn:rooij}.

\begin{lemma}
\label{lemma clique}
If $U_2=\es$ then a minimum $(C,I)$-eds $M$ of $G$  can be
found in polynomial time.
\end{lemma}

When there are no undecided vertices in the graph
we can easily find a minimum $(C,I)$-eds.
Lemma~\ref{lemma clique} tells us that clique components in the
\emph{undecided graph} $G[V\setminus(C\cup I)]$ do not
cause trouble. We use some branching rules to deal with
vertices in $U_2$.

\medskip
Consider the following simple branching rule.
For any vertex $v\in U_2$ consider two branches that either include
$v$ into the vertex cover or exclude $v$ from the vertex cover.
In the first branch we move $v$ into $C$.
In the second branch we move $v$ into $I$ and move the set
$N(v)$ of neighbors of
$v$ into $C$.

When we include a number of vertices
into the vertex cover, we reduce the parameter $p$
by the same value. Furthermore, in each branch
we move any newly-found clique component $Q$ in  $G[U_2]$  into $U_1$ and
reduce  $p$ by $|V(Q)|-1$.
The reason is that each clique has at most one vertex
that is not in the vertex cover.

Let $C(p)$ denote the worst-case running time to enumerate vertex
covers up to size $p$.
Then we have the following inequality:
\begin{equation}
\label{equation general}
C(p)\leq C(p-1-q_v)+C(p-|N(v)|-q_{N(v)}),
\end{equation}
where $q_v$ (resp., $q_{N(v)}$)
denotes the sum of $|V(Q)|-1$ over all cliques $Q$ in $G[U_2]$
that appear after removing $v$ (resp., $N(v)$) from $U_2$.

At worst, both $q_v$ and $q_{N(v)}$ are 0.
Then we end up with the recurrence
\[C(p)\leq C(p-1)+C(p-|N(v)|).\]
Note that one can always branch on vertices of degree at least $2$
in $G[U_2]$.
In this manner Fernau~\cite{kn:fernau} solves
the edge dominating set problem
in $O^*(1.6181^p)=O^*(2.6181^k)$ time which
stems from the solution of the Fibonacci recurrence
\[C(p)\leq C(p-1)+C(p-2).\]
Fomin {\em et.al.\/},~\cite{kn:fomin} refine this
as follows. Their algorithm
first branches on vertices in $G[U_2]$ of degree at least $3$
and then it considers the treewidth of the graph when all the vertices
in $G[U_2]$ have degree one or two.
If the treewidth is small the algorithm
solves the problem by dynamic programming
and if the treewidth is large the algorithm branches
further on vertices of
degree two in $G[U_2]$.
This algorithm uses exponential space and its running time depends
on the running time of the dynamic programming algorithms.

\smallskip
The method of iteratively branching on vertices of maximum degree
is powerful when this is more than two.
Unfortunately,
it seems that we can not avoid some branchings on vertices of
degree $2$,
especially when each component of $G[U_2]$ is a
$2$-path, {\em i.e.\/}, a path that consists of two edges.
We say that we are in the {\em worst case\/} when
every
component of $G[U_2]$ is a $2$-path.

\smallskip
Our algorithms branch on vertices of maximum degree and
on some other local structures in $G[U_2]$ until $G[U_2]$ has
only $2$-path components.
When we are in the worst case our algorithms deal with the graph in the
following way.
Let $P=v_0v_1v_2$ be a $2$-path in $G[U_2]$.
We say $P$ is {\em signed\/} if $v_1\in V(M)$, and
{\em unsigned\/} if $v_1\not\in V(M)$.
We use an efficient way to enumerate all signed $2$-paths
in $G[U_2]$.

\medskip
In the next section we introduce our branching rules.
\vspace{-1mm}
\section{Branching rules}
\label{sec_rules}

Besides the simple technique of branching on a vertex,
we also use the following branching rules.
Recall that in our algorithm, once a clique component $Q$ appears
in $G[U_2]$, we move $V(Q)$ into $U_1$
and reduce $p$ by $|V(Q)|-1$.

\vspace{-2mm}
\subsubsection{Tails}

Let the vertex $v_1$ have degree two.
Assume that $v_1$ has one
neighbor $v_0$ of degree one and that the other neighbor
$v_2$ has degree $>1$.
Then we call the path $v_0v_1v_2$ a {\em tail\/}.

In this paper, when we use the notation $v_0v_1v_2$ for a
tail,
we implicitly mean that the first vertex $v_0$ is the degree-$1$
vertex of the tail.
{\em Branching on a tail\/} $v_0v_1v_2$ means that we branch by
including $v_2$ into the vertex cover or excluding $v_2$
from the vertex cover.
\vspace{-1mm}
\begin{lemma}
\label{clique}
If $G[U_2]$ has a tail then we can branch with the recurrence
\begin{equation}
\label{equation 2-2}
C(p)\leq 2C(p-2) \quad\Rightarrow\quad C(p)=O(1.4143^p).
\end{equation}
\end{lemma}
\begin{proof}
Let the tail be $v_0v_1v_2$.
In the branch where $v_2$ is included into $C$, $\{v_0,v_1\}$
becomes a clique component and this is moved into $U_1$.
Then $p$ reduces by $1$ from $v_2$ and by $1$
from $\{v_0,v_1\}$.
In the branch where $v_2$ is included into $I$,
$N(v_2)$ is included into $C$.
Since $|N(v_2)|\geq 2$, $p$ also reduces by
$2$ in this branch.
\qed\end{proof}

\vspace{-6mm}
\subsubsection{$4$-Cycles}

We say that $abcd$ is a {\em $4$-cycle\/}
if there exist the four edges $ab$, $bc$, $cd$ and $da$ in the graph.
Xiao~\cite{kn:xiao} used the following lemma to obtain a
branching rule for the maximum independent set problem. In this paper
we use it for the edge dominating set problem.

\begin{lemma}
\label{4-cycle}
Let $abcd$ be a $4$-cycle in graph $G$, then any vertex
cover in $G$ contains either $a$ and $c$ or $b$ and $d$.
\end{lemma}

As our algorithm aims at finding a vertex cover,
it branches on a $4$-cycle $abcd$ in $G[U_2]$ by including
$a$ and $c$ into $C$ or including $b$ and $d$ into $C$.
Notice that we obtain the same recurrence as in Lemma~\ref{clique}.
\vspace{-3mm}
\section{A simple algorithm}
\label{sec_firsalg}
\begin{figure}[!htbp]
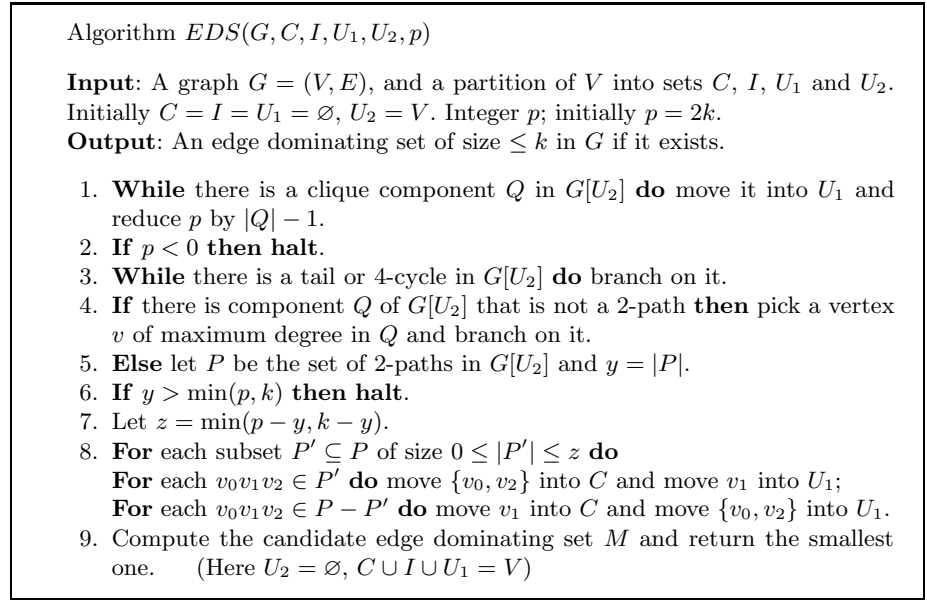
 \setbox4=\vbox{\hsize28pc
\noindent\strut
\begin{quote}
\vspace*{-5mm} Algorithm $EDS(G,C,I,U_1,U_2,p)$\\
\vspace*{-1mm}

\textbf{Input}: A graph $G=(V,E)$, and a
partition of $V$ into sets $C$, $I$, $U_1$ and $U_2$.
Initially $C=I=U_1=\es$, $U_2=V$.
Integer $p$; initially $p=2k$. \\
\textbf{Output}: An edge dominating set of size $\leq k$ in $G$
if it exists.\\

\vspace*{-2mm}
\begin{enumerate}
\item \textbf{While} there is a clique component $Q$ in $G[U_2]$
\textbf{do} move it into $U_1$ and reduce $p$ by $|Q|-1$.
\item \textbf{If} $p< 0$ \textbf{then} \textbf{halt}.
\item \textbf{While} there is a tail or $4$-cycle
in $G[U_2]$ \textbf{do} branch on it.
\item \textbf{If} there is component $Q$ of $G[U_2]$
that is not a $2$-path \textbf{then} pick a vertex $v$ of maximum degree
in $Q$ and branch on it.
\item \textbf{Else} let $P$ be the set of $2$-paths in $G[U_2]$ and $y=|P|$.
\item \textbf{If} $y>\min (p,k)$ \textbf{then} \textbf{halt}.
\item Let $z=\min(p-y, k-y)$.
\item \textbf{For} each subset $P'\subseteq P$ of
size $0\leq |P'|\leq z$ \textbf{do}

   \textbf{For} each $v_0v_1v_2\in P'$ \textbf{do}
move $\{v_0,v_2\}$ into $C$ and move $v_1$ into $U_1$;\\
   \textbf{For} each $v_0v_1v_2\in P-P'$ \textbf{do} move $v_1$
into $C$ and move $\{v_0,v_2\}$ into $U_1$.
 \item Compute the candidate edge dominating set $M$ and
return the smallest one.\ \ \ \ \
(Here $U_2=\es$, $C\cup I\cup U_1=V$)
\end{enumerate}

\end{quote} \vspace*{-5mm} \strut}  $$\boxit{\box4}$$ \vspace*{-9mm}
\caption{Algorithm $EDS(G,C,I,U_1,U_2,p)$}
\label{eds}\vspace*{-6mm}
\end{figure}

\vspace{-1mm}
Our first algorithm is described in Fig.~\ref{eds}.
The search tree consists of two parts.
First, we branch on vertices of maximum degree, tails and
$4$-cycles in Lines~3-4 until every component in $G[U_2]$
is a $2$-path. Second, we enumerate the unsigned
$2$-paths in $G[U_2]$.
In each leaf of the search tree we find an edge dominating
set in polynomial time by Lemma~\ref{lemma clique}.
We return
a smallest one.

\vspace{-6mm}
\vspace{-6mm}
\subsection{Analysis}
\label{analysis eds}

To show the correctness of the algorithm we
explain Line~6 and Line~8.

For each $2$-path in $G[U_2]$ we need at least
one edge to dominate it. So, we must have that
$y\leq p$ and $y\leq k$. This explains
the condition in Line~6.

It is also easy to see that for each unsigned $2$-path we need at
least two different edges to dominate it.
Let $p'$ be the number of unsigned $2$-paths.
In Line~8, we enumerate the
possible sets $P' \subseteq P$ of unsigned $2$-paths.
Notice that
\vspace{-1mm}
\[(y+p'\leq k \;\;\text{and}\;\; y+p'\leq p) \quad\Leftrightarrow\quad
p^{\prime} \leq z.\]

We analyze the running time of this algorithm.
Lemma~\ref{lemma clique}
guarantees that the subroutine in Line~$9$ runs in polynomial time.
We focus on the exponential part of the running time.
We prove a bound of the
size of the search tree in our algorithm with respect to measure $p$.

First, we consider the running time of Lines~3-4.

\begin{lemma}
\label{branch3}
If the graph has a vertex of degree $\geq3$ then
Algorithm $EDS$ branches with
\begin{equation}
\label{equation branch_3}
C(p)\leq C(p-1)+C(p-3) \quad\Rightarrow\quad
C(p)=O(1.4656^p).
\end{equation}
\end{lemma}
\begin{proof}
If the algorithm branches on a tail or a $4$-cycle
we have the upperbound given by (\ref{equation 2-2}).
Else the algorithm branches on a vertex of maximum degree
and generates a recurrence covered by (\ref{equation branch_3}).
Notice that (\ref{equation branch_3}) covers (\ref{equation 2-2}).
This proves the lemma.
\qed\end{proof}

\begin{lemma}
\label{pathandcycle}
If all components of the graph are paths and cycles
then the branchings of
Algorithm $EDS$ before Line~5
satisfy~(\ref{equation branch_3}).
\end{lemma}
\begin{proof}
If there is a path component of length $>2$,
then there is a tail and the algorithm branches on it
with (\ref{equation 2-2}).

If there is a component $C_l$ which is an $l$-cycle in $G[U_2]$,
the algorithm deals with it in this way:
If the cycle is a $3$-cycle, the algorithm moves it
into $U_1$ without branching since it is a clique.

If the cycle is a $4$-cycle then, according
to Lemma~\ref{4-cycle}, our algorithm
branches on it with (\ref{equation 2-2}).

If the cycle has length at least $5$,
our algorithm selects an arbitrary vertex $v_0$ and branches
on it. Subsequently it branches on the path that is created
as long as the
length of the path is greater than $2$.
When the cycle is a $5$-cycle we obtain the recurrence
\[C(p)\leq 3C(p-3) \quad\Rightarrow\quad C(p)=O(1.4423^p).\]

When the cycle is a $6$-cycle we obtain the
recurrence
\[C(p)\leq C(p-2)+C(p-3)+C(p-4) \quad\Rightarrow\quad C(p)=O(1.4656^p).\]

The two recurrences above are covered by
(\ref{equation branch_3}).
Straightforward calculations show that when the cycle
has length $\geq 7$, we also get a recurrence covered by
(\ref{equation branch_3}).
For brevity we omit the details of this analysis.
\qed\end{proof}

By Lemma~\ref{branch3} and Lemma~\ref{pathandcycle}
we know that the running time of the algorithm, before it
enters Line~5
is $O^*(1.4656^x)$, where $x$ is the size of $C$ upon
entering the loop in Line~8.
We now consider the time that is taken by the loop in Line~8
and then analyze
the overall running time.

First we derive a useful inequality.


\begin{lemma}
\label{lemma ine}
Let $r$ be a positive integer.
Then for any integer $0\leq i \leq \lfloor \frac{r}{2} \rfloor$
\begin{equation}
\label{equation ine}
{r-i \choose i}=O(1.6181^r).
\end{equation}
\end{lemma}
\begin{proof}
Notice that
\[\binom{r-i}{i} \leq \sum_{i=0}^{\lfloor \frac{r}{2} \rfloor}
\binom{r-i}{i} = F(r+1),\]
where $F(r)$ is the $r^{\mathrm{th}}$ Fibonacci number.
We have $F(r)=O(1.6181^r)$.
\qed\end{proof}

Now we are ready to analyze the running time of the algorithm.
It is clear that the loop in Line~8
takes less than $y{y \choose z}$ basic
computations.
First assume that $x\leq k$. We have that $z\leq k-y$ thus
$y+z\leq k$. If we apply Lemma~\ref{lemma ine} with $r=y+z$ we
find that
the running time of the loop in Line~8 is $O^*(1.6181^k)$.
By Lemmas~\ref{branch3} and~\ref{pathandcycle}
the running time of the algorithm is therefore bounded by
$O^*(1.4656^x \cdot 1.6181^k)=O^*(2.3715^k)$.

Assume that $x>k$. We now use that $z\leq p-y$ thus $y+z \leq p$.
By Lemma~\ref{lemma ine}
the running time of Step~8 is $O^*(1.6181^p)$.
Now $p\leq 2k-x$ and $x>k$.
The running time of the algorithm is therefore bounded by
\[O^*(1.4656^x \cdot 1.6181^p)=O^*(2.3715^k).\]

\smallskip
We summarize the result in the following theorem.

\begin{theorem}
\label{result1}
Algorithm $EDS$ solves the parameterized edge dominating
set problem in $O^*(2.3715^k)$ time and polynomial space.
\end{theorem}

\section{An improvement}
\label{sec_improved}

In this section we present an improvement
on Algorithm $EDS$.
The improved algorithm
is described in Fig.~\ref{eds1}.

The search tree of this algorithm consists of three parts.
First, we iteratively branch on vertices of degree $\geq 4$ until $G[U_2]$
has no such vertices anymore (Line~$3$).
Then we partition the vertices in $U_2$ into two parts: $V(P)$ and $U_2'$,
where $P$ is the set of $2$-path components in $G[U_2]$ and
$U_2'=U_2\setminus V(P)$.
Then the algorithm branches on vertices in $U_2'$ until $U_2'$ becomes
empty (Line~$4$-$5$).
Finally, we enumerate the number of unsigned $2$-paths
in $P$ (Line~$9$) and continues as in Algorithm $EDS$.

In Algorithm $EDS1$
a subroutine $Branch3$ deals with some components of maximum
degree $3$. It is called in Line~5. This is the major difference
with Algorithm $EDS$.
Algorithm $Branch3$ is described in Fig.~\ref{eds3d}.
The algorithm contains several simple branching cases.
They could be described in a shorter way but we
avoided doing that for analytic purposes.

\smallskip
We show the correctness of the condition in~Line~7 of
Algorithm $EDS1$.
The variable $p_0$ in Algorithm $PEDS1$ marks the decrease
of p by subroutine $Branch3$.
Note that no vertices in $V(P)$ are adjacent to vertices in $U'_2$.
Let $M_1$ be the set of edges in the solution with at least
one endpoint in $U_2'$ and let $M_2$ be the set of edges in the
solution with at least one endpoint in $V(P)$.
Then
\[M_1\cap M_2=\emptyset \quad\text{and}\quad
|M_1|+|M_2|\leq k \quad\text{and}\quad
|M_1|\geq \frac{p_0}{2}.\]
Thus $|M_2|\leq k-\frac{p_0}{2}$.
The correctness of Algorithm $EDS1$ now follows since the
only difference is the subroutine $Branch3$.

\begin{figure}[!htbp] \setbox4=\vbox{\hsize28pc
\noindent\strut
\begin{quote}
\vspace*{-5mm} Algorithm $EDS1(G,C,I,U_1,U_2,p)$\\
\vspace*{-1mm}

\textbf{Input}: A graph $G=(V,E)$ and a partition
of $V$ into sets $C$, $I$, $U_1$ and $U_2$.
Initially $C=I=U_1=\es$, $U_2=V$.
Integer $p$; initially $p=2k$. \\
\textbf{Output}: An edge dominating set of size $\leq k$ in $G$
if it exists.\\

\vspace*{-6mm}
\begin{enumerate}
\item \textbf{While} there is a clique
component $Q$ in $G[U_2]$ \textbf{do} move it into $U_1$ and
decrease $p$ by $|Q|-1$.
\item \textbf{If} $p< 0$ \textbf{halt}.
\item \textbf{While} there is a vertex $v$ of
degree $\geq 4$ in $G[U_2]$ \textbf{do}  branch on it.
\item Let $P$ denote the set of $2$-path components in $G[U_2]$
and $U_2'=U_2\setminus V(P)$. Let $y=|P|$ and $p'=p$.
\item \textbf{While} $U_2'\neq \es$ and $p\geq0$ \textbf{do}

$(G,C,I,U_1,U_2,p)=Branch3(G,C,I,U_1,U_2=U'_2\cup V(P),p)$.
\item Let $p_0=p^{\prime}-p$.

\item \textbf{If} $y> \min (p,k-p_0/2)$ \textbf{halt}.
\item Let $z=\min(p-y, k-p_0/2-y)$.
\item \textbf{For} each subset $P'\subseteq P$ of
size $0\leq |P'|\leq z$ \textbf{do}

   \textbf{for} each $v_0v_1v_2\in P'$ \textbf{do}
move $\{v_0,v_2\}$ into $C$ and move $v_1$ into $U_1$;\\
   \textbf{for} each $v_0v_1v_2\in P-P'$ \textbf{do} move $v_1$
into $C$ and move $\{v_0,v_2\}$ into $U_1$.
\item Compute the candidate edge dominating set $M$ and
return the smallest one.\ \ \ \ \
(Here $U_2=\es$, $C\cup I\cup U_1=V$.)
\end{enumerate}

\end{quote} \vspace*{-5mm} \strut}  $$\boxit{\box4}$$ \vspace*{-9mm}
\caption{Algorithm $EDS1(G,C,I,U_1,U_2,p)$}
\label{eds1}\vspace*{-6mm}
\end{figure}

\begin{figure}[!htbp] \setbox4=\vbox{\hsize28pc
\noindent\strut
\begin{quote}
\vspace*{-5mm} Algorithm $Branch3(G,C,I,U_1,U_2=U'_2\cup V(P),p)$\\

\vspace*{-2mm}
\begin{enumerate}
\item \textbf{If} there is a clique component in $G[U'_2]$ \textbf{then}
move it to $U_1$.
\item \textbf{If} there is a $2$-path component $v_0v_1v_2$ in $G[U'_2]$
\textbf{then} branch on $v_1$.
\item \textbf{If} $U'_2\neq \es$ \textbf{then}
\begin{enumerate}
\item [3.1] \textbf{If} there is a degree-$3$ vertex $v$ adjacent
to two degree-$1$ vertices in $G[U'_2]$ \textbf{then} branch on $v$.
\item [3.2] \textbf{If} there is a tail $v_0v_1v_2$ such that $v_2$ is
a degree-$2$ vertex in $G[U'_2]$ \textbf{then} branch on the tail.
\item [3.3] \textbf{If} there is a tail $v_0v_1v_2$ such
that $v_2$ is a degree-$3$ vertex in $G[U'_2]$ \textbf{then}
branch on the tail.
\item [3.4] \textbf{If} there is a degree-$3$ vertex $v$ adjacent
to one degree-$1$ vertex in $G[U'_2]$ \textbf{then} branch on $v$.
\item [3.5] \textbf{If} there is a $4$-cycle in $G[U'_2]$
\textbf{then} branch on it.
\item [3.6] \textbf{If} there is a degree-$3$ vertex $v$ adjacent
to any degree-$2$ vertex in $G[U'_2]$ \textbf{then} branch on $v$.
\item [3.7] Pick a maximum vertex $v$ in $G[U'_2]$ and branch on it.
\end{enumerate}
In addition to 3.1--3.7:
\item [*]
\textbf{If} some $2$-path component $v_0v_1v_2$ is created in 3.1 -- 3.7
\textbf{then} branch on $v_1$.
\end{enumerate}
\end{quote} \vspace*{-5mm} \strut}  $$\boxit{\box4}$$ \vspace*{-9mm}
\caption{Algorithm $Branch3(G,C,I,U_1,U_2,p)$}
\label{eds3d}\vspace*{-6mm}
\end{figure}

\subsection{Analysis of Algorithm $EDS1$}

We put the proof of the following lemma
in Appendix~\ref{analy3}.

\begin{lemma}
\label{lemma 3g}
The branchings of Algorithm $Branch3$ satisfy the recurrence
\begin{equation}
\label{equation 3g}
C(p)\leq C(p-2)+2C(p-3)
\quad\Rightarrow\quad
C(p)=O(1.5214^p).
\end{equation}
\end{lemma}

\smallskip
Algorithm $EDS1$ first
branches on vertices of
degree at least $4$.
These branchings of the algorithm satisfy
\begin{equation}
\label{equation branch_4}
C(p)\leq C(p-1)+C(p-4)
\quad\Rightarrow\quad
C(p)=O(1.3803^p).
\end{equation}

Recall that the subroutine $Branch3$ reduces $p$ by $p_0$.
The analysis without the subroutine
is similar to the analysis of Algorithm $EDS$
in Section~\ref{analysis eds}
except that $k$ is replaced by $k-\frac{p_0}{2}$ and
that Formula~(\ref{equation branch_3}) is replaced by
Formula~(\ref{equation branch_4}).
Thus without the subroutine $Branch3$ the
algorithm has a run-time proportional to
\[(1.3803 \cdot 1.6181)^{k-\frac{p_0}{2}} = 2.2335^{k-\frac{p_0}{2}}.\]

%
%
%

By Lemma~\ref{lemma 3g}
the running time of the algorithm is therefore bounded by
\[O^*(2.2335^{k-p_0/2} \cdot 1.5214^{p_0})=
O^*(2.2335^{k-p_0/2} \cdot 2.3147^{p_0/2})=O^*(2.3147^k).\]

This proves the following theorem.

\begin{theorem}
\label{theorem eds1}
Algorithm $EDS1$ solves the parameterized edge
dominating set problem in $O^*(2.3147^k)$ time and polynomial space.
\end{theorem}

\section{Kernelization}
\label{sec_kernel}

A \emph{kernelization algorithm} takes an
instance of a parameterized problem
and transforms it into
an equivalent parameterized instance
(called the \emph{kernel}), such that the new parameter is at most
the old parameter and the size
of the new instance is a function of the new parameter.

For the parameterized edge dominating set problem
Prieto~\cite{kn:prieto} presented a quadratic-time
algorithm that finds a  kernel with at most $4k^2+8k$
vertices by adapting `crown reduction techniques.'
Fernau~\cite{kn:fernau} obtained a  kernel
with at most $8k^2$ vertices.

We present a new linear-time kernelization that reduces a
parameterized edge dominating set instance $(G,k)$ to another
instance $(G',k')$ such that
\[|V(G')|\leq 2k'^2+2k' \quad\text{and}\quad
|E(G')|=O(k'^3) \quad\text{and}\quad  k'\leq k.\]

In our kernelization algorithm we first
find an arbitrary maximal matching $M_0$ in the
graph in linear time.
Let $m=|M_0|$, then we may assume that $m\geq k+1$
otherwise $M_0$ solves the problem directly.
Let
\[V_m=V(M_0) \quad\text{and}\quad V^*=V-V_m.\]
Since $M_0$ is a maximal matching, we know that $V^*$ is an
independent set.
For a vertex $v_i\in V_m$, let $x_i=|V^*\cap N(v_i)|$.
We call vertex $v_i\in V_m$ \emph{overloaded},
if $m+x_i> 2k$.
Let $A\subseteq V_m$ be the set of overloaded vertices.

\begin{lemma}
\label{overloaded}
Let $M$ be an edge dominating set $M$ of size at most $k$.
Then
\[A\subseteq V(M).\]
\end{lemma}
\begin{proof}
If an overloaded vertex $v_i \not\in V(M)$
then all neighbors of $v_i$ are in $V(M)$.
Note that at least one endpoint of each edge in $M_0$ must be
in $V(M)$ and that $V^*\cap N(v_i)$ and $V(M_0)$ are disjoint.
Therefore, $|V(M)|\geq x_i+m$. Since $v_i$ is an overloaded vertex
we have that $|V(M)|> 2k$. This implies that $|M|>k$ which is
a contradiction.
\qed\end{proof}

Lemma~\ref{overloaded}
implies that all overloaded vertices must be in the vertex
set of the edge dominating set.
We \emph{label} these vertices to indicate that these vertices
are in the vertex set of the edge dominating set.

We also label a vertex $v$ which is adjacent to
a vertex of degree one.

\smallskip
Our kernelization algorithm is presented in Fig.~\ref{kernel}.
In the algorithm
the set $A'$ denotes the set of
labeled vertices.
The correctness of the algorithm follows from the following observations.
Assume that there is a vertex $u$ only adjacent to labeled vertices.
Then we can delete it from the graph without increasing the size of
the solution.
The reason is this.
Let $ua$ be an edge that is in the edge dominating set of the original
graph where $a$ is a labeled vertex.
Then we can replace $ua$ with another edge that is
incident with $a$ to get an
edge dominating set of the new graph.
This is formulated in the reduction rule in Line~4 of the algorithm.
We add a new edge for each
labeled vertex in Line~5 to enforce that the labeled vertices are selected
in the vertex set of the edge dominating set.

\vspace*{-8mm}
\begin{figure}[!htbp] \setbox4=\vbox{\hsize28pc
\noindent\strut
\begin{quote}
\vspace*{-5mm} Algorithm $Kernel(G,k)$\\
\vspace*{-2mm}
\begin{enumerate}
\item Find a maximal matching $M_0$ in $G$.
\item Find the set $A$ of overloaded vertices and let $A'=A$.
\item If there is a vertex $v\in V_m$ that has a
degree-$1$ neighbor then delete $v$'s degree-$1$ neighbors from the
graph and let $A'\leftarrow A'\cup\{v\}$.
\item If there is a vertex $u\in V^*$ such that
$N(u)\subseteq A'$ then delete $u$ from $G$.
\item For each vertex $w\in A'$ add a new vertex
$w_i'$ and a new edge $w_i'w_i$\\
(In the analysis we assume that the new vertex $w_i'$ is in $V^*$).
\item Return $(G',k'=k)$, where $G'$ is the new graph.
\end{enumerate}
\end{quote} \vspace*{-5mm} \strut}  $$\boxit{\box4}$$ \vspace*{-9mm}
\caption{Algorithm $Kernel(G,k)$}
\label{kernel}\vspace*{-6mm}
\end{figure}

It is easy to see that each step of the algorithm can be implemented
in linear time. Therefore, the algorithm takes linear time.

\smallskip
We analyze the number of vertices in the new graph $G'$
returned by Algorithm $Kernel(G,k)$.
Note that $A'$ is a subset of $V_m$.
Let  $B=V_m-A'$.
Let $q$ be the number of edges between $V^*$ and $B$. Then

\[q=\sum_{v_i\in B} x_i\leq \sum_{v_i\in B}(2k-m)=|B|(2k-m).\]
Let
\[V^*_1=\bigcup_{v\in B} N(v)\cap V^*
\quad\text{and}\quad V^*_2=V^*-V^*_1.\]
Each vertex in $V^*_1$ is adjacent to a vertex in $B$.
Since there are at most $q$ edges between $V^*_1$ and $B$ we have
\[|V^*_1|\leq q.\]

Notice that all vertices of $V^*_2$ have only neighbors in $A'$.
In Line 4 the algorithm deletes all vertices that have
only neighbors in $A'$.
In Line 5 the algorithm adds a new vertex $v'$ and
a new edge $v'v$ for each vertex $v$ in $A'$.
Thus $V^*_2$ is the set of new vertices that are added in Line 5.
This proves
\[|V^*_2|= |A'|=2m-|B|.\]

The total number of vertices in the graph is
\[
\begin{array}{*{20}lll}
   |V_m|+|V^*_1|+|V^*_2| & \leq  & 2m +|B|(2k-m)+(2m-|B|) &\\
   {} &  =  & 4m+|B|(2k-m-1) & \\
   {} &  \leq  & 4m +2m(2k-m-1) & \quad\text{since $|B|\leq 2m$}  \\
      {} &  =  & 2m(2k-m+1) &\\
      {} &  \leq &  2k(k+1) &
\quad\text{since $m \geq k+1$.} \\
\end{array}
\]

Note that the maximal value of $2m(2k-m+1)$ as a function
of $m$ is attained for
$m=k+\frac{1}{2}$. So the function $2m(2k-m+1)$ is decreasing
for $m \geq k+1$.
\medskip

To obtain a bound for the number of edges
we partition the edge set into three disjoint sets.
\begin{enumerate}
\item Let $E_1$ be the set of edges with two endpoints in
$V_m$;
\item let $E_2$ be the set of edges between $A'$ and $V^*$, and
\item let $E_3$
be the set of edges between $B$ and $V^*$.
\end{enumerate}

It is easy to see that
\[|E_1|=O(m^2)=O(k^2) \quad\text{and}\quad |E_3|=q=|B|(2k-m)=O(k^2).\]
By the analysis above
\[|E_2|\leq |A'|\times |V^*_1|+|V^*_2|\leq |A'|q+|V^*_2| \quad\Rightarrow\quad
|E_2|=O(k^3).\]

\begin{lemma}
\label{lemma kernelresult}
Algorithm $Kernel$ runs in linear time and linear space
and it returns a kernel with at most $2k^2+2k$ vertices and $O(k^3)$ edges.
\end{lemma}

\section{Related problems}

There are standard techniques to reduce
the \emph{parameterized maximal matching problem}
that finds a maximal matching of size $k$ in a graph
to the parameterized edge dominating set problem without
increasing the input size and the parameter~\cite{kn:yannakakis}.
By Theorem~\vref{theorem eds1} we have

\begin{corollary}
\label{corollary 1}
The parameterized maximal matching problem can be
solved in $O^*(2.3147^k)$ time and polynomial space.
\end{corollary}

Another related problem is the
\emph{parameterized matrix domination problem}.
Let $M$ be an $m\times n$ matrix with entries being $0$ or $1$ and
let $k$ be an integer $k$. The problem is
to find a subset $S$ of the $1$-entries in $M$ such
that $|S|\leq k$ and every row and column of $M$ contains at
least one $1$-entry in $S$.
A parameterized matrix domination instance
reduces directly to a parameterized  edge dominating set problem
in a bipartite graph~\cite{kn:yannakakis,kn:fernau}.

\begin{corollary}
\label{corollary 2}
The parameterized matrix domination problem can be
solved in $O^*(2.3147^k)$ time and polynomial space.
\end{corollary}

%
%
%

\appendix

\section{Analysis of Algorithm $Branch3$}
\label{analy3}

In this section, we analyze Algorithm $Branch3$
presented in Fig.~\ref{eds3d} and we prove Lemma~\ref{lemma 3g}.
Initially $G[U_2']$ contains no component that is a 2-path.
We prove that in each line of Step 3, Algorithm $Branch3$
branches with  (\ref{equation 3g}), or with a better recurrence,
without leaving any newly-created 2-path components.
To be exact, some 2-path components may be created but
they are
removed immediately by an application of Line~2 in the following step.
In the analysis we merge these operations into one recurrence.
We limit the number of $2$-paths that are created in each step to
prove the upperbound on the run-time.

\begin{lemma}
\label{lemma path}
If there is a path component $P$ of length $l$ in $G[U'_2]$ then
Algorithm $Branch3$ branches with the following recurrences
until $U_2^{\prime}$ contains no more vertices of $P$.
\begin{align}
\label{equation 1-2}
C(p)&\leq C(p-1)+C(p-2)
\quad\Rightarrow \quad C(p)=O(1.6181^p)
&\quad\text{for $l=2$}\\
\label{equation 2-2_1}
C(p)&\leq C(p-2)+C(p-2)
\quad\Rightarrow\quad C(p)=O(1.4143^p)
&\quad\text{for $l=3$}\\
\label{equation 2-3}
C(p)&\leq C(p-2)+C(p-3)
\quad\Rightarrow\quad C(p)=O(1.3248^p)
&\quad\text{for $l=4$}\\
\label{equation 3-3-4}
C(p) &\leq 2C(p-3)+C(p-4)
\quad\Rightarrow\quad C(p)=O(1.3954^p)
&\quad\text{for $l=5$}\\
\label{equation 3-4-4-4}
C(p) &\leq C(p-3)+3C(p-4)
\quad\Rightarrow\quad
C(p)=O(1.4527^p)
&\quad\text{for $l=6$}\\
\label{equation 34-5}
C(p) &\leq 3C(p-4)+C(p-5)
\quad\Rightarrow\quad C(p)=O(1.3888^p)
&\quad\text{for $l=7$}\\
\label{equation 35-46}
C(p) &\leq 3C(p-5)+4C(p-6)
\quad\Rightarrow\quad
C(p)=O(1.4220^p)
&\quad\text{for $l\geq8$.}
\end{align}
\end{lemma}
\begin{proof}
Let $P$ be the path $p_0p_1\cdots p_l$.
The algorithm branches on tails of paths.
It is easy to see that Formulas~(\ref{equation 1-2}),
(\ref{equation 2-2_1})
and (\ref{equation 2-3}) hold.
When $l=5$, we first branch on $p_2$.
In the branch where $p_2$ is included into the vertex cover $C$,
we get a clique component $p_0p_1$ and a $2$-path $p_3p_4p_5$.
Then we can further reduce $p$ by at least one from $p_0p_1$ and
branch with (\ref{equation 1-2}) on $p_3p_4p_5$.
In the branch where $p_2$ is included into the independent set $I$,
$p_1$ and $p_3$ are included into $C$ and we
end up with two clique component $p_0$ and $p_4p_5$.
Then $p$ reduces further by at least one from $p_4p_5$.
Summarizing the above leads to Formula~(\ref{equation 3-3-4}).

When $l\geq6$ then, no matter whether
$p_2$ is included into the vertex cover $C$ or not,
$p$ reduces by at least two.
Then, in the first branch the algorithm branches further on an $(l-3)$-path
and in the second branch it branches further on an $(l-4)$-path.
This leads to
Formulas~(\ref{equation 3-4-4-4}) and~(\ref{equation 34-5}).

To prove Formula~(\ref{equation 35-46})
we use induction on $l$.
Assume that for all $l<l_0$ the inequality holds true.
we prove that (\ref{equation 35-46}) also
holds true for $l=l_0$, where $l_0>7$.
In the branch where $p_2$ is included into the vertex cover $C$,
the algorithm branches further on an $(l_0-3)$-path.
In the branch where $v_2$ is not included into the vertex cover,
the algorithm continues branching on an $(l_0-4)$-path.
We have that $l_0-4\geq4$. The worst recurrence
among~(\ref{equation 2-3}), (\ref{equation 3-3-4}),
(\ref{equation 3-4-4-4}), (\ref{equation 34-5}) and
(\ref{equation 35-46}) is Formula~(\ref{equation 3-4-4-4})
and the second worst recurrence is Formula~(\ref{equation 35-46}).
Furthermore, (\ref{equation 3-3-4}) is worse than (\ref{equation 34-5}).
Thus the two branches that occur after branching on $v_2$ are
bounded as follows.
\begin{enumerate}[\rm (i)]
\item In the two subbranches we further branch
with (\ref{equation 3-3-4}) and (\ref{equation 3-4-4-4})
\item in both of the two subbranches we further
branch with (\ref{equation 35-46}).
\end{enumerate}
The final recurrences created by the above two worst cases are
covered by (\ref{equation 35-46}).
This proves the claim.
\qed\end{proof}

Assume that $G[U_2^{\prime}]$ contains a component which is a
cycle $C=v_0 \ldots v_{l-1}$ of length $l$.
If the cycle is a $3$-cycle, the algorithm moves it to $U_1$
without branching since it is a clique.
If the cycle is a $4$-cycle then according to
Lemma~\ref{4-cycle} the algorithm
branches with Formula~(\ref{equation 2-2}).
If the cycle is a cycle of length at least five,
the algorithm selects a vertex and branches on it.
Subsequently, it branches on the paths created in each subbranch.
By Lemma~\ref{lemma path} we obtain the following recurrences for $C_l$.

\begin{lemma}
\label{lemma cycle}
If there is a cycle-component $C$ of length $l$
in $G[U'_2]$ the algorithm branches with the following
recurrences.
\begin{align}
\label{equation c4}
\text{$l=4$:} \;\; C(p)&\leq C(p-2)+C(p-2)
\;\Rightarrow\;\;  C(p)=O(1.4143^p)\\
\label{equation c5}
\text{$l=5$:}\;\;
C(p) &\leq 3C(p-3)
\;\Rightarrow\;\;  C(p)=O(1.4423^p)\\
\label{equation c6}
\text{$l=6$:} \;\; C(p) &\leq 2C(p-3)+2C(p-4)
\;\Rightarrow\;\;  C(p)=O(1.4946^p)\\
\label{equation c7}
\text{$l \geq 7$:} \;\; C(p) &\leq C(p-5)+6C(p-6)+4C(p-7)
\;\Rightarrow\;\; C(p)=O(1.4724^p).
\end{align}
\end{lemma}
\begin{proof}
Straightforward computations yield Formulas~(\ref{equation c4}),
(\ref{equation c5}) and (\ref{equation c6}).
We prove (\ref{equation c7}).
In the two branches
we get two paths of length $l-2$ and $l-4$.
Formulas~(\ref{equation 3-4-4-4}) and (\ref{equation 35-46})
are the two worst recurrences
among (\ref{equation 2-2_1}), (\ref{equation 2-3}),
(\ref{equation 3-3-4}), (\ref{equation 3-4-4-4}), (\ref{equation 34-5}) and
(\ref{equation 35-46}).
This
gives Formula~(\ref{equation c7}).
\qed\end{proof}

\begin{lemma}
\label{lemma step3.1}
The branching in Line~3.1 of
Algorithm $Branch3$ (together with the branching on all $2$-paths
that are created) generate
\begin{equation}
\label{equation 233}
C(p)\leq C(p-2)+2C(p-3) \quad\Rightarrow\quad
C(p)=O(1.5214^p).
\end{equation}
\end{lemma}
\begin{proof}
Assume $v$ is a degree-$3$ vertex with two
degree-$1$ neighbors in $G[U_2^{\prime}]$.
The algorithm selects $v$ and branches; either it includes $v$
into $C$
or it includes $v$ into $I$
(and adds $\{u_1,u_2,u_3\}$ to $C$).
We are interested in the number of
$2$-paths that are created in each branch.
In the first branch at most one $2$-path component is
created.
If this occurs
then the second branch creates no $2$-path.

Let the pair $(a,b)$ denote that there are $a$ $2$-paths
created in the first branch and $b$ $2$-paths
created in the second branch.
Then the possible values for $(a,b)$ are $(0,0)$, $(1,0)$, $(0,1)$
and $(0,2)$.

Once a $2$-path component is created
the algorithm branches on it.
In the first case this gives a recurrence
$C(p)\leq C(p-1)+C(p-3)$ and it leaves no $2$-path component.
In the second case
the algorithm branches with
\[C(p)\leq C(p-1-1)+C(p-1-2)+C(p-3)=C(p-2)+2C(p-3),\]
and it leaves no $2$-paths.
In the third case the algorithm branches with
\begin{eqnarray*}
C(p) &\leq& C(p-1)+C(p-3-1)+C(p-3-2)=\\
&=& C(p-1)+C(p-4)+C(p-5)\\
&\Rightarrow&
C(p)=O(1.4971^p).
\end{eqnarray*}
This case leaves no $2$-paths.
It is easy to see the above three recurrences
are covered by (\ref{equation 233}).

When the fourth case occurs
there are only three possible cases for the component that
contains $v$. We illustrate the three cases $a$, $b$ and $c$ in
Fig.~\ref{g1}.

In Case~$a$ the first branch after
deleting $v$ has a path of length $6$ and then the algorithm
branches further according to recurrence~(\ref{equation 3-4-4-4}).
In the second branch the algorithm branches further on
two $2$-paths
with the recurrence
\begin{eqnarray*}
C(p) &\leq& C(p-1-1)+C(p-1-2)+C(p-2-1)+C(p-2-2)=\\
&=& C(p-2)+2C(p-3)+C(p-4).
\end{eqnarray*}
Summarizing, we get
\[C(p)\leq C(p-4)+4C(p-5)+2C(p-6)+C(p-7) \;\;\Rightarrow\;\;
C(p)=O(1.4876^p).\]

In Case~$b$ the first branch after deleting $v$ causes the algorithm
to branch further on a degree-$3$ vertex and so on.
We obtain
the recurrence
\begin{eqnarray*}
C(p) &\leq& C(p-1-1)+C(p-1-3)+C(p-3-1)+C(p-3-2)=\\
&=&C(p-2)+2C(p-4)+C(p-5).
\end{eqnarray*}
In the second branch of Case~$b$
the algorithm branches further on two $2$-paths.
Putting these together we obtain
\[C(p)\leq C(p-3)+3C(p-5)+3C(p-6)+C(p-7) \;\;\Rightarrow\;\;
C(p)=O(1.5042^p).\]

In Case~$c$, in the first branch after deleting $v$
the algorithm branches on a degree-$3$ vertex and so on.
This yields
\begin{eqnarray*}
C(p)&\leq& C(p-1-2)+C(p-1-2)+C(p-3-1)+C(p-3-2)=\\
&=&2C(p-3)+C(p-4)+C(p-5).
\end{eqnarray*}
In the second branch of Case~$c$
the algorithm branches on two $2$-paths.
If we take them together we get
\begin{eqnarray*}
C(p)&\leq& 2C(p-4)+2C(p-5)+3C(p-6)+C(p-7) \\
&\Rightarrow&
C(p)=O(1.4941^p).
\end{eqnarray*}

Since the solution of (\ref{equation 233}) satisfies
$C(p)=O(1.5214^p)$,
it follows that (\ref{equation 233}) covers all the cases.

This proves the lemma.
\qed\end{proof}


{\begin{figure}[h]
\begin{center}
\includegraphics[width=1\textwidth]{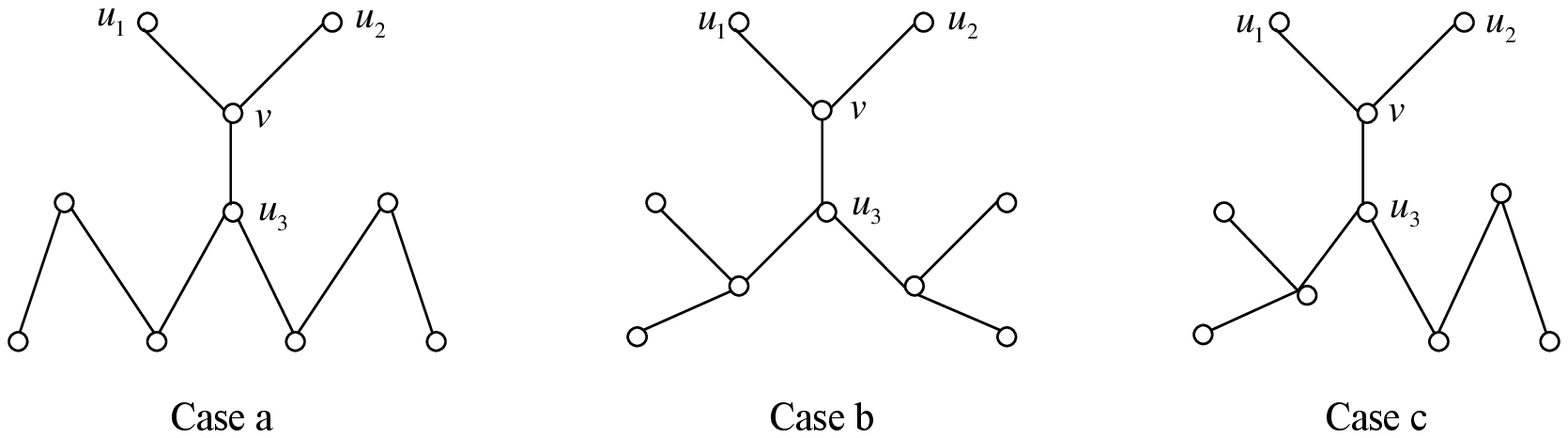}
\end{center}
\caption{The three possible graphs for
Case~$(0,2)$}\label{g1}
\end{figure}}

\begin{lemma}
\label{lemma step3.2}
In Line~3.2 of Algorithm $Branch3$ the algorithm branches
with
\[C(p)\leq C(p-2)+2C(p-3).\]
\end{lemma}
\begin{proof}
Assume that $v_0v_1v_2$ is the tail and $v_2$ is
a degree-$2$ vertex in $G[U'_2]$.
The algorithm branches on $v_2$ by including it into $C$
or including it into $I$ (and including its neighbors into $C$).

We consider the number of $2$-paths that are created in each branch.
If the component that contains the tail is a $5$-path,
then the algorithm branches on it
according to Recurrence~(\ref{equation 3-3-4}).
Otherwise, it is impossible to create a $2$-path component
after removing $v_2$.

There are at most two $2$-path components created in the second branch
since there is no degree-$3$ vertex adjacent to two degree-$1$ vertices.
If only one $2$-path component is created,
the algorithm branches according to
\begin{eqnarray*}
C(p) &\leq& C(p-2)+C(p-2-1)+C(p-2-2)=\\
&=& C(p-2)+C(p-3)+C(p-4).
\end{eqnarray*}
If two $2$-path components are created
the algorithm branches with
\begin{eqnarray*}
C(p) &\leq& C(p-2)+C(p-2-2)+2C(p-2-3)+C(p-2-4)=\\
&=& C(p-2)+C(p-4)+2C(p-5)+C(p-6).
\end{eqnarray*}
All the recurrences above are weaker than $C(p)\leq C(p-2)+2C(p-3)$.
This proves the lemma.
\qed\end{proof}

\begin{lemma}
\label{lemma step3.3}
In Line~3.3 of Algorithm $Branch3$ the algorithm branches with
\[C(p)\leq C(p-2)+2C(p-3).\]
\end{lemma}
\begin{proof}
The proof of Lemma~\ref{lemma step3.3} is
similar to the proof of Lemma~\ref{lemma step3.4}.
\qed\end{proof}

\begin{lemma}
\label{lemma step3.4}
In Line~3.4 of Algorithm $Branch3$
the algorithm branches with
\[C(p)\leq C(p-2)+2C(p-3).\]
\end{lemma}
\begin{proof}
Assume that $v$ is a degree-$3$ vertex having one degree-$1$ neighbor
in $G[U'_2]$.
The algorithm branches on $v$ by including it into $C$ or
including it into $I$.
Since Line~3.1 and Line~3.2 do no longer apply we can simply assume
that in $G[U'_2]$ there is no more degree-$3$ vertex adjacent to
two degree-$1$ vertices nor any tail $v_0v_1v_2$ with $v_2$ being
a degree-$2$ vertex.

Under this assumption
we analyze the number of $2$-path components
created in each branch.

Let $u_0u_1u_2$ be a $2$-path created after
removing $v$ (or $N(v)$).
Then there are at least two edges
between $\{u_0,u_1,u_2\}$ and $v$ (or $N(v)$)
otherwise,  before branching on $v$
the condition of Line~3.1 or Line~3.2 holds.
This implies that, after removing $v$, at most one $2$-path
component is created.

If a $2$-path component is created in the first branch
then no $2$-path component is created in the other branch.
Therefore the branching of the algorithm satisfies
\begin{eqnarray*}
C(p) &\leq& C(p-1-1)+C(p-1-2)+C(p-3)=\\
&=& C(p-2)+2C(p-3).
\end{eqnarray*}

If no $2$-path component is created in the first branch
then at most two $2$-path components are created
in the second branch.
In the worst case we first branch on the degree-$3$ vertex $v$ and
then branch on two $2$-path components in the second branch with
\[C(p)\leq C(p-2)+2C(p-3)+C(p-4).\] Therefore, we get
\begin{equation}
\label{equation b3b22}
C(p)\leq C(p-1)+C(p-5)+2C(p-6)+C(p-7) \;\;\Rightarrow\;\;
C(p)=O(1.5181^p).
\end{equation}
It follows that the worst recurrence is $C(p)\leq C(p-2)+2C(p-3)$.
\qed\end{proof}

\begin{lemma}
\label{lemma step3.5}
In Line~3.5 of Algorithm $Branch3$ the algorithm branches with
\[C(p)\leq C(p-2)+2C(p-3).\]
\end{lemma}
\begin{proof}
Let $v_1v_2v_3v_4$ be a $4$-cycle in $G[U'_2]$,
the algorithm branches by including $v_1$ and $v_3$ or by
including $v_2$ and $v_4$
into the vertex cover.
Note that after Line~3.5
there are no more degree-$1$ vertices in $G[U'_2]$.
Thus in each branch at most one $2$-path component is created.
Therefore, the algorithm branches first
with $C(p)\leq 2C(p-2)$ for the $4$-cycle and then
it possibly branches further on a $2$-path in each branch.
This leads to the recurrence
\begin{eqnarray*}
C(p) &\leq& C(p-2-1)+C(p-2-2)+C(p-2-1)+C(p-2-2)=\\
&=& 2C(p-3)+2C(p-4)\\
&\Rightarrow&
C(p)=O(1.4946^p).
\end{eqnarray*}
This is covered by $C(p)\leq C(p-2)+2C(p-3)$.
\qed\end{proof}

\begin{lemma}
\label{lemma step3.6}
In Line~3.6 of Algorithm $Branch3$ the algorithm branches with
\[C(p)\leq C(p-2)+2C(p-3).\]
\end{lemma}
\begin{proof}
Assume that $v$ is a degree-$3$ vertex adjacent to at least
one degree-$2$ neighbor in $G[U'_2]$. The algorithm branches
on $v$ by including it into $G$ or $I$.

In the first branch no $2$-path is created
otherwise there is a $4$-cycle in $G[U'_2]$ before
the branching on $v$.

In the second branch, where $N(v)$ is moved into $C$,
there are at most two $2$-path components created.
Note that for any $2$-path component $u_0u_1u_2$ that is created
after removing $N(v)$ there are at least two edges between
$\{u_0,u_1,u_2\}$ and  $N(v)$, and at least one vertex in
$N(v)$ is a degree-$2$ vertex.
Therefore, we have (\ref{equation b3b22}) as an upperbound.
\qed\end{proof}

\medskip
Now we are ready to complete the proof of Lemma~\ref{lemma 3g}.

\begin{proof}
Notice that
Lemmas~\ref{lemma step3.1}, \ref{lemma step3.2}, \ref{lemma step3.3},
\ref{lemma step3.4}, \ref{lemma step3.5} and \ref{lemma step3.6}
guarantee that, if any of the Lines~3.1 - 3.6 are called,
the algorithm branches according to Formula~(\ref{equation 3g}).

In Line~3.7 the induced subgraph $G[U'_2]$ has only two kinds of
components: each component is
either a cycle or a $3$-regular graph without any $4$-cycle.
Lemma~\ref{lemma cycle} proves that the branching on a
cycle gives a recurrence which is no worse
than~(\ref{equation 3g}).

Now we may assume that there are only $3$-regular components.
In this case the algorithm selects an arbitrary vertex $v$ and
branches on it.
According to the analysis in the proof of Lemma~\ref{lemma step3.6}
no $2$-path components are created
after removing $v$.
In the branch where $N(v)$ is removed at most one $2$-path
is created.
Note that for any $2$-path component
$u_0u_1u_2$, created after removing $N(v)$,
there are five edges between $\{u_0,u_1,u_2\}$ and  $N(v)$,
because each vertex is a degree-$3$ vertex before the branching.
In the worst case the algorithm still branches with
\begin{eqnarray*}
C(p) & \leq & C(p-1)+C(p-3-1)+C(p-3-2)=\\
&=& C(p-1)+C(p-4)+C(p-5).
\end{eqnarray*}
This is weaker than Formula~(\ref{equation 3g}).

Therefore, the branching of Algorithm $Branch3$
satisfies~(\ref{equation 3g}).
\qed\end{proof}

\end{document}